\documentclass{svproc}
\usepackage[pdftex]{graphicx}
\usepackage{url}
\usepackage{hyperref}

\begin{document}
\mainmatter              % start of a contribution
\title{Four Things People Should Know About Migraines}
\titlerunning{Four Things People Should Know  About Migraines}  % abbreviated title (for running head)
\author{Mohammad S. Parsa \and Lukasz Golab*}
\authorrunning{Parsa and Golab} % abbreviated author list (for running head)
\tocauthor{Mohammad S. Parsa, Lukasz Golab}
\institute{University of Waterloo, Waterloo, Ontario, Canada N2L 3G1\\
\email{\{msparsa,lgolab\}@uwaterloo.ca}}

\maketitle              % typeset the title of the contribution

\begin{abstract}
Migraine literacy among the public is known to be low, and this lack of understanding has a negative impact on migraineurs' quality of life.  To understand this impact, we use text mining methods to study migraine discussion on the Reddit social media platform.  We summarize the findings in the form of "four things people should know about chronic migraines": it is a serious disease that affects people of all ages, it can be triggered by many different factors, it affects women more than men, and it can get worse in combination with the COVID-19 virus.
\end{abstract}

\section{Introduction}
Despite being a common cause of disability worldwide \cite{Steiner2018}, chronic migraines are underdiagnosed, undertreated, and not well understood \cite{Patwardhan2006,Minen2016,Viana2020}. Additionally, a recent study reports that migraine literacy is significantly lower among non-migraineurs \cite{Goodhew2019}. In this paper, we analyze social media discussions on the Reddit platform to understand how this lack of understanding affects migraineurs’ quality of life (QoL). We summarize the findings in the form of “four things people should know about migraines.” 

The closest work to ours surveyed 33 patients and found that their QoL improved as their families learned more about migraines \cite{RuizdeVelasco2003}. This study and others \cite{Minen2016,Viana2020} also reported that the QoL of migraineurs is negatively impacted by a lack of understanding among healthcare providers. As an example of how misinformation can spread, Saffi et al. \cite{Saffi2020} examined YouTube videos about migraines and found that frequently viewed videos contain non-evidence-based information and were not created by healthcare professionals. Our goal is to go one step further and create a list of information items to increase migraine literacy among the public. Furthermore, online social media platforms such as Facebook \cite{Egan2016} and Twitter \cite{Deng2020,Linnman2013,Nascimento2014,Zhang2020} were recently used to study migraines. These studies, however, investigated migraineurs’ online behaviour and self-presentation rather than public understanding.  

The social media platform we study, Reddit, has over 430 million (anonymous) users as of October 2021 \cite{Kemp2021}, participating in over 180,000 user-created discussion communities referred to as subreddits. Each subreddit focuses on a specific topic and is named after the topic: e.g., r/babybumps hosts discussions about pregnancies. A discussion begins when a user submits an initial post; others then submit comments in response. Unlike Twitter’s limit of 280 characters per message, Reddit does not limit the length of the posted content, making it an ideal platform for a text-mining study to assess migraine literacy among the public and the effect of misunderstandings on migraineurs’ lives.

We study Reddit discussions containing the term “migraine” that were published between September 2015 and September 2021. First, we identify the subreddits with the most such discussions, to understand the contexts in which these discussions take place. Second, we apply a topic modelling algorithm to segment the discussions into topics.  Finally, we organize the discovered topics into themes, and, from the themes, we identify four aspects of migraines that, if understood, could make migraineurs’ lives easier.

\section{Methods}

Using Reddit’s official Application Programming Interface (API)\footnote{https://www.reddit.com/dev/api/}, we downloaded all 722,902 submissions (posts and comments) containing the word “migraine” between September 2015 and September 2021. These posts and comments are authored by 317,874 distinct Reddit users. Although users are not required to disclose their location or demographics, previous work reports that the Reddit user base is skewed toward English-speaking North Americans under 50 years of age \cite{Similarweb2022,PewResearchCenter}. 

Each post and comment includes the corresponding text as well as the name of the subreddit in which it appeared. The methods we used to analyze the data are summarized in Figure 1 and discussed below.

\begin{figure}[t]
  \centering
  \includegraphics[width=\linewidth]{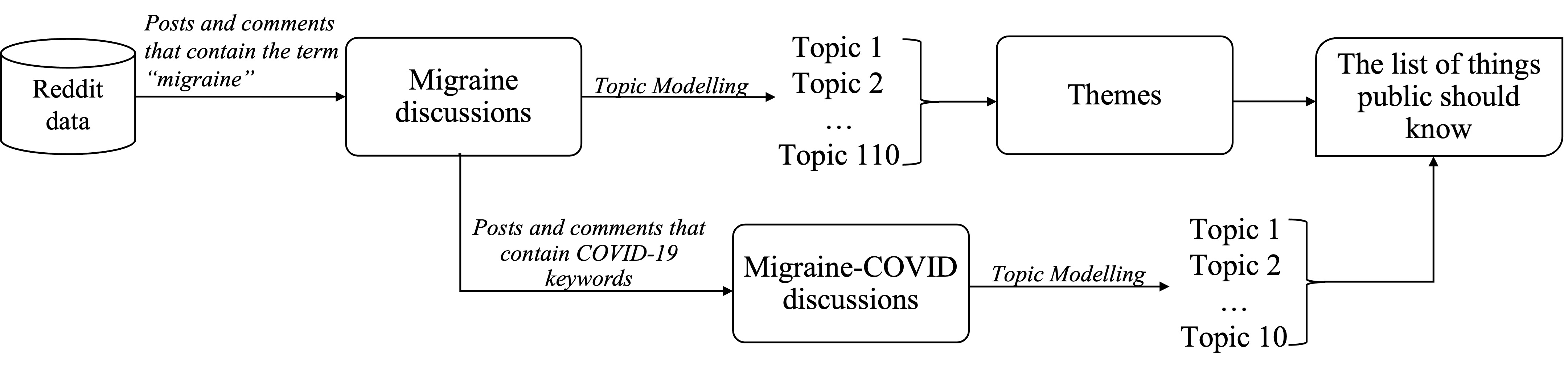}
  \caption{Summary of methods}
  \label{fig:methods}
\end{figure}

We begin by identifying the subreddits with the most submissions that contain the word “migraine”. The names of these subreddits suggest the context in which migraines are discussed. For example, migraine posts on r/babybumps are likely to be related to migraines experienced during pregnancies.

Next, before applying topic modelling, we perform standard text preprocessing \cite{Deng2020,Maier2018,Curiskis2020} to remove words that are not semantically meaningful. For each post, we remove stopwords (i.e., words that only serve a grammatical purpose, such as “and” or “the”), links to Web pages, and common words that appear in more than 10\% of posts containing the word “migraine” (such as “migraine” and “headache”). 

We then apply topic modelling on the (posts and comments with the word “migraine” in the) above subreddits. Topic modelling is a text mining technique that segments documents based on words used. Documents assigned to the same segment use similar words and thus they are likely to discuss the same topic. We use a state-of-the-art topic modelling algorithm that considers word semantics during segmentation (i.e., different words that have similar meanings, such as “dog” and “puppy”, are considered similar) \cite{Curiskis2020}. We implemented this algorithm in Python based on its description in \cite{Curiskis2020} and personal communication with the authors of \cite{Curiskis2020} to clarify how to choose an optimal number of topics.  Our code is publicly available on \url{https://github.com/msparsa/doc2vec-topic-modelling}. 

The output of the topic modelling algorithm consists of the following, for each topic: the top ten words that best describe the topic, 20 word bigrams and trigrams\footnote{Bigrams and trigrams are sequences of two or three consecutive words appearing in the posts, such as “strong migraine” or “having a migraine”.}  that most frequently appear in posts and comments assigned to that topic, and 1\% of the posts and comments that best represent the topic. After reading the output for each topic, we manually group similar topics into themes.  Each author did so independently, followed by a discussion to resolve the differences. The inter-annotator agreement was 79\%. Finally, from the discovered themes, we identify four aspects of migraines that, if known by the public, could make migraineurs’ lives easier.

%COVID-19 Discussions

Recent studies suggest that the COVID-19 pandemic created new challenges for migraineurs \cite{Cerami2021,Consonni2021}. We thus also perform a dedicated topic modelling analysis on discussions containing the word “migraine” as well as at least one word from the list of COVID-19 related words frequently used in social media, as identified in \cite{Chen2020}. Examples of such words include “coronavirus”, “covid” and “pandemic”.

\section{Results}

Table 1 lists the 20 subreddits with the most posts and comments containing the word “migraine”. These subreddits together contain nearly half of such posts and comments, with the remainder scattered across 1062 other subreddits. The second column in Table 1 includes a summary of the topics found in each subreddit (due to space constraints, we omit a detailed discussion of the topic modelling output).
 
\begin{table}[h!]
    \centering
    \caption{Subreddits with the most migraine discussions, including a summary of topics for each subreddit, the number of posts and comments containing the word “migraine”, and the number of distinct users who authored these posts and comments}
    \label{tab:subreddits}
    \begin{tabular}{p{0.27\linewidth}|p{0.5\linewidth}|p{0.07\linewidth}|p{0.07\linewidth}|p{0.07\linewidth}}
    Subreddit&A summary of topic modeling results&Users&Posts&Comments\\ \hline
         migraine&Discussions about migraine triggers, treatments, and symptoms.&28382&23102&125541\\ \hline
AskReddit&Questions about migraine symptoms, triggers, and medications.&51712&7663&80441\\ \hline
AskDocs&Questions about migraine symptoms and treatments.&9989&3054&7558\\ \hline
AmItheA**hole&Migraineurs discussing situations in which they may have overreacted, and non-migraineurs discussion situations in which they may have acted inappropriately towards migraineurs.&4738&1309&6306\\ \hline
ChronicPain&Migraineurs discussing pain and medications.&2458&914&5381\\ \hline
todayilearned&Users discussing new things they learned about migraine symptoms, medications, and triggers.&4267&913&5046\\ \hline
AskWomen&Discussions about migraine triggers specific to women (i.e., menstruation and birth control pills), as well as relationship advice&2890&903&4808\\ \hline
unpopularopinion&Migraineurs pointing out that people are not aware of how painful migraines are. &1290&886&4790\\ \hline
keto&Discussions about the Ketogenic diet and its effects on migraines.&2613&805&4425\\ \hline
birthcontrol&Discussions about methods that do not trigger migraines, such as Intrauterine devices.&2510&748&4152\\ \hline
explainlikeimfive&Discussions of migraine symptoms, medications, and triggers.&3416&731&3791\\ \hline
BabyBumps&Migraineurs sharing their experiences during pregnancies, e.g., more frequent migraine attacks. &2367&726&3525\\ \hline
Fibromyalgia&Similar to r/ChronicPain, discussions about migraine treatments and symptoms.&1793&703&3422\\ \hline
relationships&Migraineurs discussing relationship issues with friends, family, and partners.  &2838&667&3362\\ \hline
TwoXChromosomes&Discussions similar to r/AskWomen.&2693&620&3342\\ \hline
IAmA&People introduce themselves as “I am a (doctor, teacher, etc.)” and offer to answer questions. Migraineurs asked doctors about migraine medications and triggers.&2864&617&3320\\ \hline
twentyonepilots&Subreddit for fans of the band twenty one pilots, one of whose albums is titled Migraine&2060&590&3068\\ \hline
trees&Discussions about cannabis as a pain reliever and as a migraine trigger.&2577&543&3005\\ \hline
RocketLeagueExchange&A subreddit for trading Rocket league trophies. One of the trophies is called migraine.&740&538&2869\\ \hline
childfree&Migraineurs expressing fears of migraine attacks while pregnant and fears of not being good parents.&1872&528&2710\\ \hline
    \end{tabular}
    
\end{table}

The top subreddit is r/migraine, a community for people suffering from migraines, followed by two question-answering subreddits: r/AskReddit (general questions) and r/AskDocs (medical questions). Other top subreddits include r/AmItheA**hole (people describing situations in which they may have acted inappropriately), and communities related to family and relationship issues (r/birthcontrol, r/BabyBumps, r/relationships and r/childfree). Note that there are two “false positives”: r/RocketLeagueExchange and r/twentyonepilots. The former is a community of players of the computer game Rocket League Exchange, in which a migraine is a trophy. The latter is a community of fans of the band Twenty One Pilots, one of whose albums is titled “Migraine”. Excluding these two false positives, the total number of topics is 130.

When grouping together the 130 topics from the above 18 subreddits (i.e., not including the two false positives), we identified seven themes: symptoms, treatments, triggers, work and school issues, relationship issues, women’s issues, and mental health issues. Figure 2 shows the percentage of topics we assigned to each theme, and Figure 3 shows the percentage of topics from each theme for each subreddit. We summarize the seven themes below. 

\begin{figure}[t]
  \centering
  \includegraphics[width=0.75\linewidth]{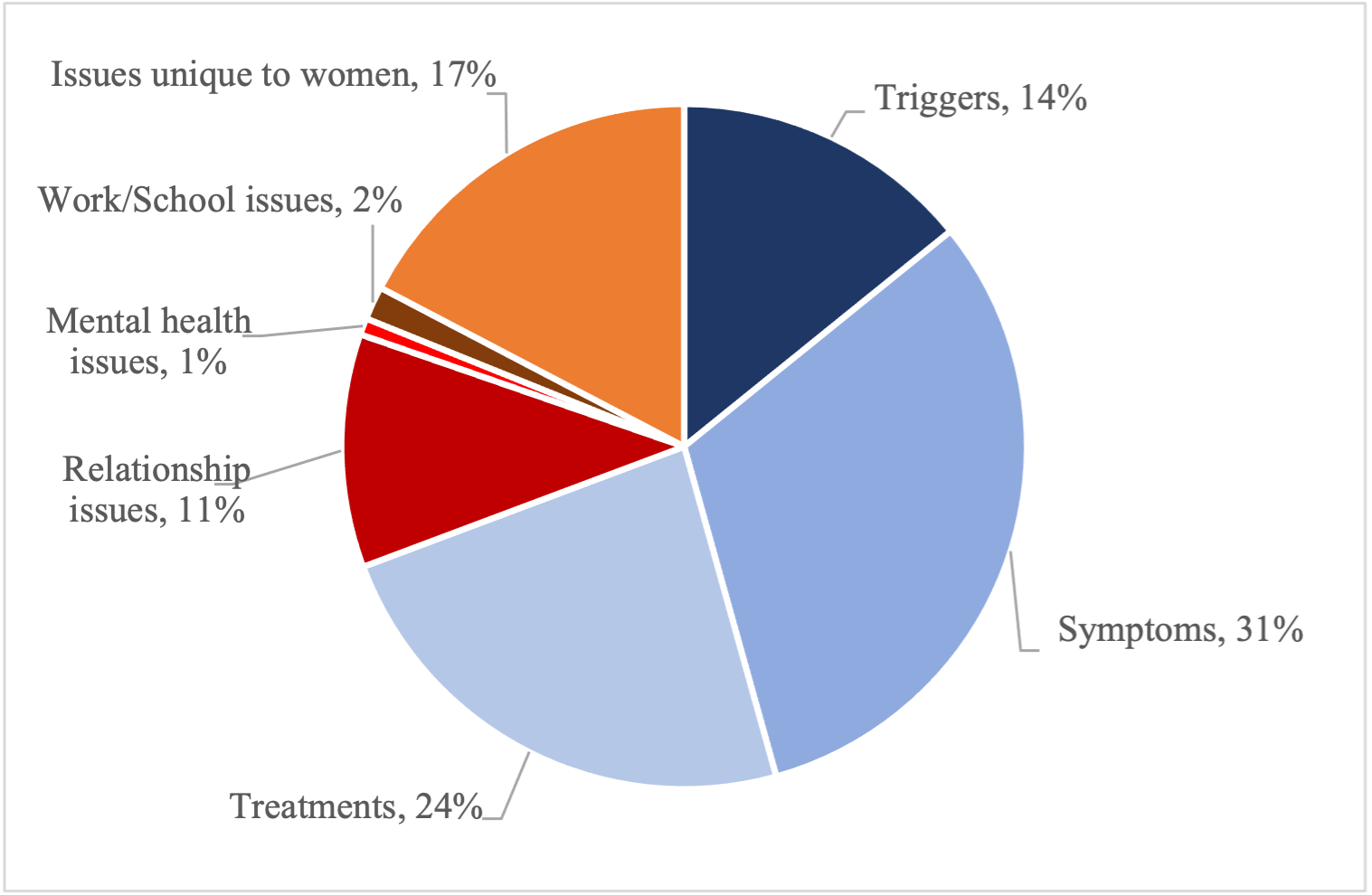}
  \caption{Distribution of migraine-related discussions }
  \label{fig:methods}
\end{figure}

\begin{figure}[t]
  \centering
  \includegraphics[width=\linewidth]{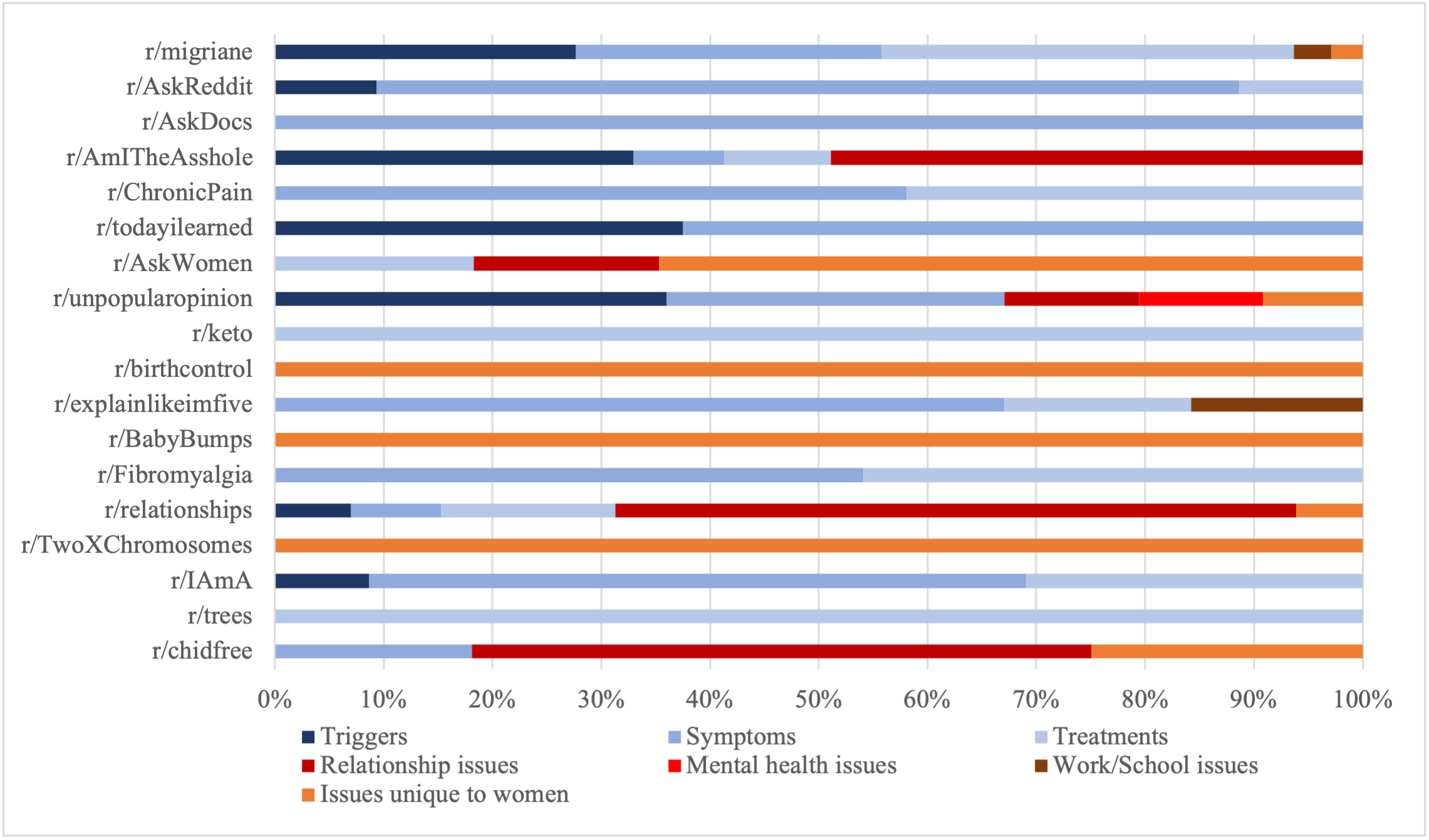}
  \caption{Distribution of migraine-related discussions on each subreddit}
  \label{fig:methods}
\end{figure} 

\textbf{Symptoms:} Migraineurs shared their symptoms to express frustration, and seek help and emotional support. Symptoms discussed include muscle pain (e.g., neck pain), temporary blindness in one or both eyes, nausea, and auras. Additionally, non-migraineurs asked about these symptoms to understand whether they have the disease and thus should seek medical support. Moreover, some of these users were worried that severe headaches could be caused by a brain tumour. 

\textbf{Treatments:} Discussions pointed out that different treatments are effective for different people. Frequently discussed medications include Botulinum toxin (Botox), Excedrin, Topiramate (Topamax), Ibuprofen, Sumatriptan (Imitrex), and Paracetamol.

\textbf{Triggers:} Frequently discussed triggers include stress, lack of sleep, caffeine, dairy, alcohol, bright lights, loud noises, smoke, perfume, Monosodium glutamate (MSG), artificial sweeteners, weather changes, menstruation, and changes in estrogen levels (e.g., using hormonal birth control pills). Some migraineurs mentioned caffeine and cannabis smoke as triggers while others found these to be pain relievers. Others commented on the usefulness of mobile apps such as “Migraine Buddy” to keep track of triggers and migraine attacks.

\textbf{Work/School issues:} Migraineurs discussed their inability to work or study during a migraine attack, earning poor grades as a result. There were also some discussions about employers refusing to grant sick days due to migraines.

\textbf{Relationship issues:} Discussions on r/relationships and r/AmItheA**hole revealed that migraineurs experience conflicts with family members, friends, partners, and the public in general.  Migraineurs reported being criticized for not performing household and parenting duties, not attending gatherings with family or friends, and overreacting in situations involving bright lights or noise (such as yelling at a crying baby). Migraineurs often cited poor migraine literacy as a reason for these conflicts. On the other hand, people in a relationship with a migraineur complain that it is difficult to relate to their partner, with some believing that having a migraine is not a good excuse to evade responsibility. 

\textbf{Issues unique to women:} Discussions on subreddits related to women and pregnancies revealed that migraines impact the lives of women, especially those in a relationship, in two ways. First, women report more frequent and painful migraine attacks during menstruation and pregnancy. These attacks also affect women’s sexual lives and lead to relationship problems with their partners. Second, parents with migraines reported that having a child creates more stress, and that crying babies can trigger migraine attacks. For these reasons, some migraineurs choose a child-free lifestyle and use non-hormonal birth control methods, such as Intrauterine devices (IUD).   

\textbf{Mental health issues:} The issues discussed above (e.g., relationship issues and academic difficulties) negatively impact the mental wellbeing of migraineurs. Topic modelling revealed that these issues contribute to stress, anxiety, loneliness, and depression in migraineurs, which lead to more frequent attacks. Additionally, migraineurs report being unsatisfied with their lives, believing the disease prevents them from achieving their life goals.

%Migraine-covid discussions

Finally, we discuss the topic modelling results using posts and comments containing the word “migraine” and at least one word related to the COVID-19 pandemic. We found a total of 8437 such posts and comments, leading to ten topics grouped into three themes. 
First, some users with migraines experienced more frequent and more painful attacks when sick with COVID-19 and after getting the vaccine, as well as fears that a migraine may be a symptom of being infected with the virus. Second, some users report that wearing a facial covering can trigger migraine attacks. Third, migraineurs reported negative impacts of lockdowns on their mental health: becoming even more isolated from family and friends, and relationship troubles due to spending more time at home with family or roommates (e.g., arguing over loud music or bright lights).

\section{Four Things People Should Know About Migraines}

Based on topic modelling and  grouping of topics into seven themes, we suggest the following information items to inform the public about migraines. We believe the selected information items can address common misunderstandings about migraines. Below we include quotes from Reddit discussions as examples of misunderstandings and their effects on migraineurs’ lives. 

\textbf{1. It is a serious disease that affects people of all ages.} 
\%Chronic migraines affect people of all ages. 
Symptoms include head and muscle pain, temporary blindness, and nausea, making it difficult to do basic tasks, let alone perform well in school or in a workplace. Co-workers, workplace supervisors, teachers, family members and relationship counsellors should keep in mind that what appears as laziness or lack of productivity is a side effect of this serious condition. Migraineurs would like to live a normal life but they cannot, and being perceived as lazy only makes them feel worse.
\\

\textit{“Yes! I get bad migraine and I am just told ‘not to think about it’ and that everyone gets headache so suck it up.”}
\\

\textit{“I skipped class the other day because I had a migraine. It just so happens that I also had homework due that day, and according to my friends who were in class, my teacher kept mentioning how ‘convenient’ it was, and that I just hadn’t done the homework so was avoiding her”}
\\

\textit{“my wife felt a migraine coming on … So I called my mother and asked her if we could reschedule for next weekend … A couple hours later my Dad texts me saying that it was disrespectful to cancel with such little notice”}
\\

\textit{“This is how life working with a migraine should be for us all.  3 out of the 11 of my coworkers have chronic migraines.  Working there means, if I’m in a lot of pain, people actually understand.  First place I’ve EVER worked like that.”}
\\

\textit{“My parents always want me to talk about my problems, but they somehow make it to where it’s my fault. `You have a migraine? Maybe you shouldn’t \_\_\_\_\_\_\_' is their most common response.”}

\textbf{2. Various factors can trigger a migraine attack, and these factors differ from one person to another.}
\%A migraine attack can be triggered by many things, including 
Potential triggers include stress, weather changes, menstruation, light, noise, and strong smells. Even if something is not a trigger for one person, it can be for another.
\\

\textit{“For me, it’s bananas. If I eat bananas, I get a migraine”}
\\

\textit{“As someone who gets migraines and suffers from chemical sensitivity, it’s shitty when people act like it’s fake or not serious when they douse themselves in perfume”}
\\

\textit{“I only learned recently that caffeine can be a trigger because it’s always helped me. When I have a migraine the first thing I do is drink some coffee or a diet mountain dew or something.”}

\textbf{3. Migraine affects more women than men,}
\%Chronic migraines are a serious women’s health issue, 
with consequences on relationships and family planning, including the choice of birth control methods. Additionally, migraines may get worse during menstruation or pregnancy.
\\

\textit{“My first pregnancy I had migraines consistently until 7 weeks and then I didn’t have another until 35 weeks. At 35 weeks, I had a bad one that lasted about a week and brought my BP up to the point where I needed a half day of monitoring at Labor \& Delivery.”}
\\

\textit{“My migraines are often linked to my menstrual cycle too. I’ve had the Mirena IUD (hormonal not copper) for a few years because my gyno told me that many birth control pills can worsen migraine auras.”}

\textbf{4. The COVID-19 pandemic has created additional hardships for migraineurs}
\%The COVID-19 virus is a concern for migraineurs 
since it can make their headaches worse. Another source of stress is that it is difficult to distinguish between a “usual” headache and a headache that may be a symptom of the virus. 
\\

\textit{“I had to try 4 different styles of masks to find one that didn’t give me a migraine”}
\\

\textit{“I had a bad case of long covid last year which resulted in dysautonomia and research and my doctors confirmed that my aimovig (anti-CGRP) which I was on for migraine made my respiratory and cardiac symptoms far worse”}

\section{Conclusions}

In this paper, we leveraged the Reddit social media platform to characterize online discussions about migraines. Text mining using topic modelling, combined with manual inspection, allowed us to produce a list of four things the public should know about chronic migraines. While most of this information, such as the severity of migraines \cite{Mannix2016,Sokolovic2013}, variety of triggers \cite{Dahlof1995,Hauge2010TriggerAura} and prevalence in women \cite{Burch2020}, is known to healthcare providers, our social media analysis revealed that the public may not be aware of these issues, and this creates additional stress and hardships for migraineurs.  

One limitation of this study is its focus on English language content on Reddit, motivating future research on migraine literacy among non-English speakers. Furthermore, a natural next step is to investigate the best way to leverage social media again, this time to disseminate information about migraines. 
Another interesting direction is to examine the role of mobile apps such as “Migraine Buddy” in helping migraineurs to keep track of symptoms and helping healthcare providers to find effective treatments.

\bibliographystyle{abbrv}
\bibliography{references}

\begin{thebibliography}{10}

\bibitem{Burch2020}
R.~Burch.
\newblock {Epidemiology and Treatment of Menstrual Migraine and Migraine During Pregnancy and Lactation: A Narrative Review}.
\newblock {\em Headache}, 60(1):200--216, 2020.

\bibitem{Cerami2021}
C.~Cerami, C.~Crespi, S.~Bottiroli, G.~C. Santi, G.~Sances, M.~Allena, T.~Vecchi, and C.~Tassorelli.
\newblock {High perceived isolation and reduced social support affect headache impact levels in migraine after the Covid-19 outbreak: A cross sectional survey on chronic and episodic patients.}
\newblock {\em Cephalalgia}, 41(14):1437--1446, 2021.

\bibitem{Chen2020}
E.~Chen, K.~Lerman, and E.~Ferrara.
\newblock {Tracking Social Media Discourse About the COVID-19 Pandemic: Development of a Public Coronavirus Twitter Data Set}.
\newblock {\em arXiv}, 6(2):e19273, 2020.

\bibitem{Consonni2021}
M.~Consonni, A.~Telesca, L.~Grazzi, D.~Cazzato, and G.~Lauria.
\newblock {Life with chronic pain during COVID-19 lockdown: the case of patients with small fibre neuropathy and chronic migraine}.
\newblock {\em Neurological Sciences}, 42(2):389--397, 2021.

\bibitem{Curiskis2020}
S.~A. Curiskis, B.~Drake, T.~R. Osborn, and P.~J. Kennedy.
\newblock {An evaluation of document clustering and topic modelling in two online social networks: Twitter and Reddit}.
\newblock {\em Information Processing and Management}, 57(2):102034, 2020.

\bibitem{Dahlof1995}
C.~G. Dahl{\"{o}}f and E.~Dimen{\"{a}}s.
\newblock {Migraine patients experience poorer subjective well-being/quality of life even between attacks}.
\newblock {\em Cephalalgia}, 15(1):31--36, 1995.

\bibitem{Deng2020}
H.~Deng, Q.~Wang, D.~P. Turner, K.~E. Sexton, S.~M. Burns, M.~Eikermann, D.~Liu, D.~Cheng, and T.~T. Houle.
\newblock {Sentiment analysis of real-world migraine tweets for population research}.
\newblock {\em Cephalalgia Reports}, 3:1--9, 2020.

\bibitem{Egan2016}
K.~G. Egan, J.~S. Israel, R.~Ghasemzadeh, and A.~M. Afifi.
\newblock {Evaluation of migraine surgery outcomes through social media}.
\newblock {\em Plastic and Reconstructive Surgery - Global Open}, 4(10):1084, 2016.

\bibitem{Goodhew2019}
S.~C. Goodhew.
\newblock {Migraine Literacy and Treatment in a University Sample}.
\newblock {\em SN Comprehensive Clinical Medicine}, 1(10):749--757, 2019.

\bibitem{Hauge2010TriggerAura}
A.~W. Hauge, M.~Kirchmann, and J.~Olesen.
\newblock {Trigger factors in migraine with aura}.
\newblock {\em Cephalalgia}, 30(3):346--353, 2010.

\bibitem{Kemp2021}
S.~Kemp.
\newblock {Digital 2021 October Global Statshot Report — DataReportal – Global Digital Insights}, 2021.

\bibitem{Linnman2013}
C.~Linnman, N.~Maleki, L.~Becerra, and D.~Borsook.
\newblock {Migraine Tweets – What can online behavior tell us about disease?}
\newblock {\em Cephalalgia}, 33(1):68--69, 2013.

\bibitem{Maier2018}
D.~Maier, A.~Waldherr, P.~Miltner, G.~Wiedemann, A.~Niekler, A.~Keinert, B.~Pfetsch, G.~Heyer, U.~Reber, T.~H{\"{a}}ussler, H.~Schmid-Petri, and S.~Adam.
\newblock {Applying LDA Topic Modeling in Communication Research: Toward a Valid and Reliable Methodology}.
\newblock {\em Communication Methods and Measures}, 12(2):93--118, 2018.

\bibitem{Mannix2016}
S.~Mannix, A.~Skalicky, D.~C. Buse, P.~Desai, S.~Sapra, B.~Ortmeier, K.~Widnell, and A.~Hareendran.
\newblock {Measuring the impact of migraine for evaluating outcomes of preventive treatments for migraine headaches}.
\newblock {\em Health and Quality of Life Outcomes}, 14(1):1--11, 2016.

\bibitem{Minen2016}
M.~T. Minen, E.~Loder, L.~Tishler, and D.~Silbersweig.
\newblock {Migraine diagnosis and treatment: A knowledge and needs assessment among primary care providers}.
\newblock {\em Cephalalgia}, 36(4):358--370, 2016.

\bibitem{Nascimento2014}
T.~D. Nascimento, M.~F. DosSantos, T.~Danciu, M.~DeBoer, H.~van Holsbeeck, S.~R. Lucas, C.~Aiello, L.~Khatib, M.~A. Bender, J.-K. Zubieta, and A.~F. DaSilva.
\newblock {Real-Time Sharing and Expression of Migraine Headache Suffering on Twitter: A Cross-Sectional Infodemiology Study}.
\newblock {\em Journal of Medical Internet Research}, 16(4):e96, 2014.

\bibitem{Patwardhan2006}
M.~B. Patwardhan, G.~P. Samsa, R.~B. Lipton, and D.~B. Matchar.
\newblock {Changing physician knowledge, attitudes, and beliefs about migraine: Evaluation of a new educational intervention}.
\newblock {\em Headache}, 46(5):732--741, 2006.

\bibitem{PewResearchCenter}
{Pew Research Center}.
\newblock {Social Platform Use by Demographic, April 2021}.
\newblock https://www.marketingcharts.com/wp-content/uploads/2021/04/Pew-Social-Platform-Use-by-Demographic-Apr2021.png. Accessed: 2022-02-04.

\bibitem{RuizdeVelasco2003}
I.~Ruiz~de Velasco, N.~Gonz{\'{a}}lez, Y.~Etxeberria, and J.~Garcia-Monco.
\newblock {Quality of Life in Migraine Patients: A Qualitative Study}.
\newblock {\em Cephalalgia}, 23(9):892--900, 2003.

\bibitem{Saffi2020}
H.~Saffi, T.~P. Do, J.~M. Hansen, D.~W. Dodick, and M.~Ashina.
\newblock {The migraine landscape on YouTube: A review of YouTube as a source of information on migraine}.
\newblock {\em Cephalalgia}, 40(12):1363--1369, 2020.

\bibitem{Similarweb2022}
{Similarweb}.
\newblock {Reddit.com Traffic Analytics {\&} Market Share}, {January} 2020.
\newblock https://www.similarweb.com/website/google.com/traffic. Accessed: 2022-02-05.

\bibitem{Sokolovic2013}
E.~Sokolovic, F.~Riederer, T.~Szucs, R.~Agosti, and P.~S. S{\'{a}}ndor.
\newblock {Self-reported headache among the employees of a Swiss university hospital: prevalence, disability, current treatment, and economic impact}.
\newblock {\em The journal of headache and pain}, 14(1):29, 2013.

\bibitem{Steiner2018}
T.~J. Steiner, L.~J. Stovner, T.~Vos, R.~Jensen, and Z.~Katsarava.
\newblock {Migraine is first cause of disability in under 50s: will health politicians now take notice?}
\newblock {\em The journal of headache and pain}, 19(1):17, 2018.

\bibitem{Viana2020}
M.~Viana, F.~Khaliq, C.~Zecca, M.~D. Figuerola, G.~Sances, V.~Di~Piero, B.~Petolicchio, M.~Alessiani, P.~Geppetti, C.~Lupi, S.~Benemei, R.~Iannacchero, F.~Maggioni, M.~E. Jurno, S.~Odobescu, E.~Chiriac, A.~Marfil, F.~Brighina, N.~Barrientos~Uribe, C.~P{\'{e}}rez~Lago, C.~Bordini, F.~Lucchese, V.~Maffey, G.~Nappi, G.~Sandrini, and C.~Tassorelli.
\newblock {Poor patient awareness and frequent misdiagnosis of migraine: findings from a large transcontinental cohort}.
\newblock {\em European Journal of Neurology}, 27(3):536--541, 2020.

\bibitem{Zhang2020}
P.~Zhang and S.~Bhaskarabhatla.
\newblock {How advocacy affects Twitter migraine conversations: A pilot cross-sectional survey of Northeast American “migraine” landscape on Twitter from May to June 2020}.
\newblock {\em Cephalalgia Reports}, 3:1--16, 2020.

\end{thebibliography}

\end{document}